# Superconducting Nanocircuits for Topologically Protected Qubits

Sergey Gladchenko [a], David Olaya [a], Eva Dupont-Ferrier [a], Benoit Douçot [b], Lev B. Ioffe [a], and Michael E. Gershenson [a]

[a] Department of Physics and Astronomy, Rutgers University,

136 Frelinghuysen Rd., Piscataway, NJ 08854, USA

[b] Laboratoire de Physique Théorique et Hautes Energies, CNRS UMR 7589,

Universités Paris 6 et 7, 4 place Jussieu, 75005 Paris, France

**For successful realization of a quantum computer, its building blocks (qubits) should be simultaneously scalable and sufficiently protected from environmental noise. Recently, a novel approach to the protection of superconducting qubits has been proposed. The idea is to prevent errors at the "hardware" level, by building a fault-free (topologically protected) logical qubit from "faulty" physical qubits with properly engineered interactions between them. It has been predicted that the decoupling of a protected logical qubit from local noises would grow exponentially with the number of physical qubits. Here we report on the proof-of-concept experiments with a prototype device which consists of twelve physical qubits made of nanoscale Josephson junctions. We observed that due to properly tuned quantum fluctuations, this qubit is protected against magnetic flux variations well beyond linear order, in agreement with theoretical predictions. These results demonstrate the feasibility of topologically protected superconducting qubits.**

For implementation of quantum correction codes, the decoherence time of a qubit, $\tau_d$, should be at least $10^4$ times longer than the time of a single operation, $\tau_0$ [1,2]. For the realization of a large ratio $\tau_d/\tau_0$, several requirements should be simultaneously satisfied. The decoherence rate is controlled by two processes: the transitions between the states "0" and "1" of a qubit, which usually involves energy relaxation, and the fluctuations of the relative phase between these states. For reduction of the energy relaxation rate, the energy difference between the states "0" and "1", $\Delta_{01}$, should be small (for a schematic energy diagram of a qubit, see Fig. 1): this reduces the probability of emission of photons, phonons and other excitations [3]. For reduction of dephasing, the qubit should be designed in such a way that $\Delta_{01}$, which controls the phase difference $\sim \int \Delta_{01} dt$ between "0" and "1" states, would be unaffected by uncontrollable changes in the qubit environment ("noise"). Note that a small value of $\Delta_{01}$ is also expected to be less susceptible to the fluctuations of the physical quantity that sets this energy scale. Finally, in order to reduce the operation time $\tau_0$, the gap $\Delta_{12}$ that separates the logical states of the qubit from the rest of its spectrum should be large ($\tau_0 < h/\Delta_{12}$).

It is believed that the physical sources of noise acting upon qubits are local, which makes the task of large-scale quantum computation realistic [4]. Suppression of noises by improving materials involved in the qubit fabrication is a difficult (though possible [5,6]) task. Alternatively, one might decouple a qubit from local noises. One approach to such a decoupling is based on tuning the qubit control parameters in such a way that the qubit becomes less susceptible to noise. The existence of such a "sweet spot" in the qubit parameters space, where the qubit is decoupled from noise in linear order, has been already established for superconducting qubits [7,8,9,10]. At the same time, these experiments indicate that the linear-order decoupling is currently insufficient for running long quantum computations. An alternative approach is based on the error prevention at the "hardware" level [11]. It has been proposed that a fault-free logical qubit can be build from "faulty" physical qubits if the states of the logical qubit are protected by nontrivial symmetries which emerge for properly engineered interactions between physical qubits [12,13,14,15]. Implementation of this idea would offer decoupling from local noises well beyond linear order, even in a relatively small device. In the present work, we describe the proof-of-concept experiments with prototypes of topologically protected qubits that demonstrate viability of this approach.

In this work, "faulty" physical qubits are represented by Josephson elements with an effective Josephson energy $V_R \approx E_{2R} \cos(2\phi)$ where $\phi$ is the phase difference across each element (these elements are shown as "rhombi" on Figs. 1a,b and 2a)). Each element is implemented as a superconducting loop interrupted by four Josephson junctions (JJs) and threaded by the magnetic flux $\Phi_R = \Phi_0/2$ ($\Phi_0$ is the superconducting flux quantum) [16,17,18]. An individual JJ is characterized by the Josephson energy $E_J$ and the charging energy $E_C = e^2/2C$ where $C$ is the tunnel junction capacitance [24]. If $E_C$ is negligible, the rhombus has two degenerate classical states (see Fig. 1a) which correspond to the phase difference $\Delta\phi = \pm\pi/2$ across the rhombus (the supercurrents circulate clock-wise or counter-clock-wise in the rhombus); these states are separated by the energy barrier $2E_{2R} \approx 4(\sqrt{2}-1)E_J$. A non-zero (but small) charging energy causes rare tunneling events between these states and removes the degeneracy. As a result, the states "0" and "1" of such a qubit, $\frac{1}{\sqrt{2}}\left(\left|-\frac{\pi}{2}\right\rangle + \left|\frac{\pi}{2}\right\rangle\right)$ and $\frac{1}{\sqrt{2}}\left(\left|-\frac{\pi}{2}\right\rangle - \left|\frac{\pi}{2}\right\rangle\right)$, are separated by a small energy gap [13,18]

$$t \approx E_J^{3/4} E_C^{1/4} \exp\left(-1.6\sqrt{E_J/E_C}\right), \quad (1)$$

schematically shown as splitting of red lines in Fig. 1a.

A single rhombus is unprotected against local noises: by deforming the potential $V_R(\phi)$ (Fig. 1a), noise induces fluctuations of the energy difference between $-\pi/2$ and $\pi/2$ states and, thus, dephasing. This deformation can be represented by an additional term in the qubit energy, $V_R \approx E_{2R}\cos(2\phi) + E_{1R}\sin(\phi)$; the second term describes both the effect of noise (the time-dependent part of $E_{1R}$) and an asymmetry of the rhombus (the time-independent part of $E_{1R}$). The $E_{1R}$ term lifts the degeneracy of the classical states of a faulty qubit. Topological protection a logical qubit implies that though its energy is described by a similar expression, $V(\varphi) \approx -E_2\cos(2\varphi) - E_1\cos(\varphi)$ (see below), the ratio $E_1/E_2$ is greatly reduced in comparison with a single physical qubit. The idea of our experiment is to demonstrate that $E_1/E_2$ is indeed strongly suppressed in a logical qubit consisting of just a few physical qubits. The transport in a system characterized by the energy $V(\varphi) \approx -E_2\cos(2\varphi)$ is mediated by the objects with charge 4e (correlated pairs of Cooper pairs) [13,17,18]. The absence of the $E_{1R}$ term signifies the localization of single Cooper pairs. In the charge basis, two states of a logical qubit differ by the parity of the number of Cooper pairs (the $Z_2$ group).

In its simplest form, a protected logical qubit consists of a chain of "$\cos(2\phi)$" Josephson elements which connects a superconducting "island" to a current lead with a large capacitance $C$ to the ground (Fig. 1b). A large value of $C$ suppresses the phase fluctuation in the lead (the phase of the lead, $\varphi_A$, is chosen 0 in Fig. 1b). The qubit logical variable is the phase of the superconducting "island" $\varphi$, i.e. the sum of the phase differences across individual rhombi ($\varphi = 0, \pi$ for an even number of rhombi in a chain). The double periodicity of Josephson energies of individual rhombi leads to the double periodicity of the chain energy $V(\varphi) = -E_2\cos(2\varphi)$ (Fig. 1c). In the quasiclassical limit ($C_0 \to \infty$), the sates $|0\rangle$ and $|\pi\rangle$ are degenerate, they are separated by an energy barrier $E_2 \approx \frac{\pi^2}{2}\frac{E_{2R}}{N}$ [19]. A finite probability of tunneling between the sates $|0\rangle$ and $|\pi\rangle$ removes the degeneracy and results in a finite energy splitting $\Delta_{01}$ between the global logical states of the qubit, $\frac{1}{\sqrt{2}}(|0\rangle \pm |\pi\rangle)$.

The non-local global logical states of the rhombi chain are symmetry-protected from local noises. Energy relaxation in the chain (i.e. transitions between states $|0\rangle$ and $|\pi\rangle$) is suppressed because a single rhombus cannot flip its phase by $\pi$ - it costs too much energy. However, a pair of rhombi can flip simultaneously, and these flips induced by quantum fluctuations help to suppress dephasing. Indeed, because the $\pm\pi/2$ states of an individual rhombus enter the energies of the global states of the chain symmetrically, the effect of the local noise is averaged out by the fluctuation-induced flips. For example, the global states in a two-rhombi chain (Fig. 1b) are $|0\rangle = \frac{1}{\sqrt{2}}\left(\left|-\frac{\pi}{2},-\frac{\pi}{2}\right\rangle + \left|\frac{\pi}{2},\frac{\pi}{2}\right\rangle\right)$ and $|1\rangle = \frac{1}{\sqrt{2}}\left(\left|-\frac{\pi}{2},\frac{\pi}{2}\right\rangle + \left|\frac{\pi}{2},-\frac{\pi}{2}\right\rangle\right)$. The local noise changes the energy difference between $\left|\frac{\pi}{2}\right\rangle$ and $\left|-\frac{\pi}{2}\right\rangle$ states of the first and second rhombi by $\delta E_1$ and $\delta E_2$, respectively. The noise might change the energies of the global states, but this change is the same for both global states in the first order of the perturbation theory. The difference in energies of the global states appears only in the second order: $\delta E_{01} = \frac{\delta E_1 \delta E_2}{\Delta_{12}}$. In a longer chain which consists of $N$ physical qubits, the effect of a local noise is predicted to be suppressed up to the $N^{th}$ order of the perturbation theory due to this mechanism [13].

Quantum fluctuations help to establish a global coherent state across the chain: the stronger the quantum fluctuations of the phase differences across individual rhombi in the chain, the better the decoupling from local noises. On the other hand, excessively strong quantum fluctuations suppress the Josephson energy barrier, $E_2$, between the logical states of a chain, which might lead to flips of the end phase φ (this process can be regarded as a half-vortex crossing of the chain). Note that an increase of the number of rhombi in a chain has a similar effect: it improves the protection from noise, but decreases the phase stiffness between the ends of the chain. These contradictory requirements can be reconciled by optimizing the qubit design and fine-tuning the parameters of individual Josephson junctions, $E_J$ and $E_C$. For example, the energy $E_2$ can be increased by connecting several chains in parallel and making the island capacitance $C_0$ larger (Fig. 1d). For a moderate number of chains ($\leq 4$), realization of a significant value of $E_2$ requires $E_J/E_C \geq 3$, whereas a sizable $\Delta_{12}$ is achieved for $E_J/E_C < 6$ (see Fig. 4).

The reported experiments were designed to test the key theoretical prediction that even a relatively small prototype logical qubit with properly engineered interactions between its "faulty" elements is protected against variations of external parameters well beyond the linear order. For example, the effect of the magnetic flux noise on the logical states of the $N$-element chain can be expressed in terms of the deviations of the flux in each rhombus from its optimum value, $\delta\Phi_i = \Phi_i - \Phi_0/2$, as follows [13,15]:

$$\frac{\delta\Phi_{eff}}{\Phi} \frac{E_J}{\Delta_{12}} = \prod_{i=1}^{N} \frac{\delta\Phi_i}{\Phi_0} \frac{E_J}{\Delta_{12}} . \quad (2)$$

Note that the product in Eq. 2 reflects the fact that it is sufficient to have $\delta\Phi_i = 0$ in a single rhombus in order to induce $\delta\Phi_{eff} = 0$ and, thus, the $\cos(2\varphi)$-periodic energy of the whole chain. Equation 2 shows that for an efficient noise decoupling, the energy gap $\Delta_{12}$ between the logical states and the first excited state of the qubit should be large (as it was mentioned above, this is also required for a decrease of the operation time $\tau_0 \sim \hbar/\Delta_{12}$). The decoupling from other types of local noise is described by similar expressions with $E_J$ being replaced by a relevant energy scale (e.g., $E_C$ for the charge noise); this decoupling is typically more efficient because these energies are smaller than $E_J$. Our experimental test exploits the fact that Eq. 2 describes the decoupling of a qubit from both ***time-dependent*** flux noises and ***time-independent*** flux variations. In the experiment with a prototype device which consists of three $N = 4$ chains connected in parallel, we have observed dramatic suppression of the effect of flux variations over a substantial flux range near $\Phi_R = \Phi_0/2$. This indicates that (a) the scattering of junction parameters and individual rhombi areas can be made sufficiently small for the realization of symmetry-protected superconducting qubits, and (b) our theoretical model captures all essential features of real devices.

The design of a prototype device is shown in Fig. 2. The devices were fabricated using multi-angle electron-gun deposition of Al films through a nanoscale lift-off mask. According to our numerical calculations, significant noise protection is realized if the values of $E_J/E_C$ for all Josephson junctions in the rhombi are within ~30%. To reduce the scattering of JJ parameters, we have developed a qubit design in which all Josephson junctions are formed by the aluminum strips of a well-controlled width intersecting at the right angle (for details, see Supplementary Information). Because the experiments were conducted at $T \sim 30\text{-}50$ mK; the requirement $T \ll \Delta_{12} < 0.5 E_C$ (see Fig. 4) implies that $E_C$ should be at least 0.5 - 1K. Taking into account that the specific capacitance of $Al_2O_3$-based tunnel barriers is $\sim 50$ fF/$\mu m^2$ [20], the latter condition implies sub-micron (0.1-0.2 µm) in-plane dimensions of individual junctions. We have tested several devices with the width of Al strips $W = 0.15$ - $0.18$ µm, normal-state resistance of JJs in the rhombi $R_N = 2.4$-$5$ k$\Omega$, and the ratio $E_J/E_C = 2$-$5$ (see Table 1). In order to realize sizable values of both $E_2$ and $\Delta_{12}$, three $N = 4$ chains were connected in parallel and the central strip was shared by all the chains.

For the measurements of the $V(\varphi_{AB})$ dependence, the rhombi array was included in a superconducting loop of a large area ($\sim 110$ µm$^2$) interrupted by two larger (0.3 µm×0.3 µm) JJs. In order to suppress quantum fluctuations in these larger JJs, the SQUID-like device was shunted by an inter-digital capacitor $C \sim 3 \cdot 10^{-14}$ F. The phase difference $\varphi_{AB}$ across the array was controlled by varying the magnetic flux $\Phi_L$ in the SQUID loop. Because of a factor-of-100 difference between the areas of individual rhombi and the SQUID loop, the phase difference across the chains can be significantly varied without affecting the phase difference across individual rhombi. For protection from external high-frequency noise and non-equilibrium quasiparticles, this SQUID-like device was flanked by two meander-type inductances $L$. The thickness $t$ of the horizontal Al strips in the meanders shown in Fig. 2b is a factor of 2-3 smaller than that in the vertical strips. As a result, the superconducting gap varies significantly within each meander's segment (the critical temperature in Al films substantially depends on their thickness in the range of ~10-30 nm). It is expected that the gap variations in the meanders "trap" non-equilibrium quasiparticles generated outside of the device [21,22]. The charging effects in the device were probed by applying the gate voltage $V_g$ to the gate capacitor $C_g \sim 2 \cdot 10^{-18}$ F connected to the central strip common for all chains.

The $V(\varphi_{AB})$ dependence for the rhombi array was studied by measuring the probability of switching of the device into the resistive state by current pulses for different values of the external magnetic field $B$ (for details, see Supplementary Materials). Figure 3 shows the dependence of the switching current $I_{SW}$ (defined as the current pulse amplitude that causes switching of the device into the resistive state with probability 0.5) on the magnetic field for a device with the ratio $E_J/E_C = 2.7$. The range of the magnetic field in Fig. 3 corresponds to the magnetic flux through a single rhombus, $\Phi_R$, varying between $-\Phi_0/2$ and $\Phi_0/2$. The switching current oscillates with the magnetic flux $\Phi_L$ through the large loop of the SQUID-type device. The period of oscillations, $\Delta\Phi_L$, depends on $\Phi_R$: $\Delta\Phi_L = \Phi_0$ for all values of $\Phi_R$ except for $\Phi_R \approx (n+1/2)\Phi_0$, where the period is halved ($\Delta\Phi_L = \Phi_0/2$). Note that in the classical regime ($E_J/E_C > 20$), such a period halving has already been observed [23]. The beatings of oscillations with $\Delta\Phi_L = \Phi_0$, observed at $\Phi_R/\Phi_0 = \pm 1/8, \pm 1/4$, and $\pm 3/8$, are due to the flux quantization in the intermediate-size loops between adjacent rhombi chains (the area of these loops is four times greater than the area of a single rhombus). Below we focus on the regime $\Phi_R \approx \pm\frac{1}{2}\Phi_0$ where the states $\left|\frac{\pi}{2}\right\rangle$ and $\left|-\frac{\pi}{2}\right\rangle$ of individual rhombi are almost degenerate and the rhombi array is expected to be protected from noise.

The oscillations of $I_{SW}$ with the period $\Delta\Phi_L = \Phi_0$ vanish near $\Phi_R \approx \pm\frac{1}{2}\Phi_0$ (Fig. 3c). In this regime, the effective Josephson energy of a rhombus, $V_R = E_{2R}\cos(2\pi\Phi_R/\Phi_0)$, becomes small and the supercurrent of single Cooper pairs is blocked by quantum fluctuations. Observation of the oscillations of $I_{SW}$ with the period $\Delta\Phi_L = \Phi_0/2$ suggests that the supercurrent is carried by correlated pairs of Cooper pairs with charge 4e [13,17,18]. Comparison between the results obtained for the 4x3 rhombi array (Fig. 3c) and a single two-rhombi chain (Fig. 3d) shows that the oscillations with the period $\Delta\Phi_L = \Phi_0$ and, thus, the term $E_1 \cos(\varphi)$ in the qubit energy, are suppressed in the array much more efficiently over a relatively wide range of magnetic fields near $\Phi_R = \Phi_0/2$. These results are in good agreement with Eq. 2: indeed, if the suppression is due to quantum fluctuation, the term $\cos(\varphi_{AB})$ responsible for the oscillations with the period $\Delta\Phi_L = \Phi_0$ should appear in the energy of an $N = 4$ chain only in the 4$^{th}$ order in flux deviations from the optimal values $\Phi_R \approx (n+1/2) \Phi_0$. The presence of the second harmonic of the critical current oscillations together with the vanishing of the first harmonic provides an essential test for the proper strength of quantum fluctuations in the JJ rhombi. This observation also indicates that the scattering of JJ parameters in the studied devices is relatively small: only the (axially)-symmetric rhombi contribute to the order of protection $N$.

The height of the energy barrier which separates the states of the rhombi array, $2E_2$, can be found from the amplitude $I_2 = \dfrac{4eE_2}{\hbar}$ of the oscillations of switching current with period $\Delta\Phi_L = \Phi_0/2$. Figure 4 shows that the values of $2E_2/E_C$ measured for the devices with different values of $E_J/E_C$ are in good agreement with our numerical simulations (for simulation details, see Supplementary Materials). The observed agreement verifies the validity of theoretical assumptions which have been also used in the calculations of $\Delta_{12}$ (see Fig. 4). The gap $\Delta_{12}$ is the smallest of two gaps: one is associated with the excitations inside the chain with $\varphi$ being fixed, and another one – with the fluctuations of $\varphi$ around its classical value. In the quasiclassical case ($E_J/E_C \gg 1$), $\Delta_{12}$ coincides with the former gap $t_{2R} \sim t^2/E_J$ (see Eq. 1), which is smaller than the latter gap ($\sim \sqrt{32E_2 E_C}$) [14]. However, for the realization of a sizable value of $\Delta_{12}$, the ratio $E_J/E_C$ should not be too large, and the numerical simulations are required beyond applicability of quasiclassical approximation (see Supplemental Materials). Figure 4 shows that the aforementioned criteria for a proper qubit operation can be satisfied within the optimal range $E_J/E_C \sim 3\text{-}6$.

Another probe of quantum fluctuations in the studied array is provided by the measurements of the effect of the gate voltage on the switching current. In the absence of quantum fluctuations, the critical current of the device coincides with the critical current of three rhombi chains connecting the central strip to one of the leads. Quantum fluctuations, which result in tunnelling of the phase of the central strip between 0 and $\pi$, reduce $I_2$. The offset charge $\Delta q = V_g/C_g$ induced by the gate modulates the phase of the tunnelling amplitudes $t_{isl} \to t_{isl} e^{i\Delta q \Delta \varphi}$, affecting the interference of processes with $\Delta\varphi = \pm\pi$. Thus, in presence of quantum fluctuations, one might expect to observe modulation of $I_2$ by $V_g$. Indeed, Figs. 5ab show that for the device with $E_J/E_C = 4.7$, the switching probability oscillates with the gate voltage $V_g$. The amplitude of oscillations, $\Delta I_{SW}$, in the regime $\Phi_R \approx \Phi_0/2$ is in good agreement with our calculations of the dependence $\Delta I_2(E_J/E_C)$ (note that no fitting parameters are involved in this comparison). The period of oscillations, which corresponds to charging of the central strip with charge $2e$, is approximately the same in both regimes, $\Phi_R = 0$ (Fig. 5a) and $\Phi_R = \Phi/2$ (Fig. 5ab). This is expected for relatively long (1 ms) current pulses used in these measurements: though the transport of single Cooper pairs is suppressed by quantum fluctuations in this regime, still there is a considerable probability of tunnelling of a Cooper pair to/from the island over a long time scale.

The reported results indicate that the topological protection can be realized in a Josephson circuit with a properly tuned ratio $E_J/E_C$. The rhombi array studied in this work can be used as a qubit protected from local noise in the fourth order if one of the leads is replaced with an island which has a relatively small capacitance to the ground. The phase of this island (0 or π) would be the logical variable of such a qubit. It is expected that the ratio $\tau_d/\tau_0$ for such a qubit, due to the protection from local noises, would be much greater than that for an unprotected superconducting qubit. For the full characterization of the protected qubit, we plan to measure the energy $\Delta_{12}$ (~$h/\tau_0$) by direct spectroscopic measurements and to study the dephasing time by observing Rabi oscillations. Further reduction of the JJ dimensions (i.e. larger $E_C$), with simultaneous increase of the transparency of the tunnel barrier (i.e. larger $E_J$), will allow for an increase of the operational temperature of protected qubits. Finally, it is worth emphasizing that our work reports on the first observation of the coherent transport of pairs of Cooper pairs in a small-size rhombi array in the quantum regime, combined with the absence of conventional Cooper pair coherence. The persistence of this phenomenon in larger arrays would imply the appearance of a new thermodynamic phase characterized by $\langle\cos 2\varphi\rangle \neq 0$ with $\langle\cos\varphi\rangle = 0$, $Z_2$ topological order parameter which can be regarded as "superconductor nematic" [13,14,15].

Table 1

| Device | Area (μm$^2$) | $R_N$ (kΩ) | $E_C$ (K) | $E_J/E_C$ |
|--------|---------------|------------|-----------|-----------|
| 1 | 0.165×0.165 | 4.78 | 0.68 | 2.2 |
| 2 | 0.153×0.153 | 3.27 | 0.79 | 2.7 |
| 3 | 0.150×0.180 | 2.82 | 0.69 | 3.7 |
| 4 | 0.173×0.173 | 2.43 | 0.62 | 4.7 |
| 5 | 0.180×0.180 | 2.49 | 0.57 | 5.0 |

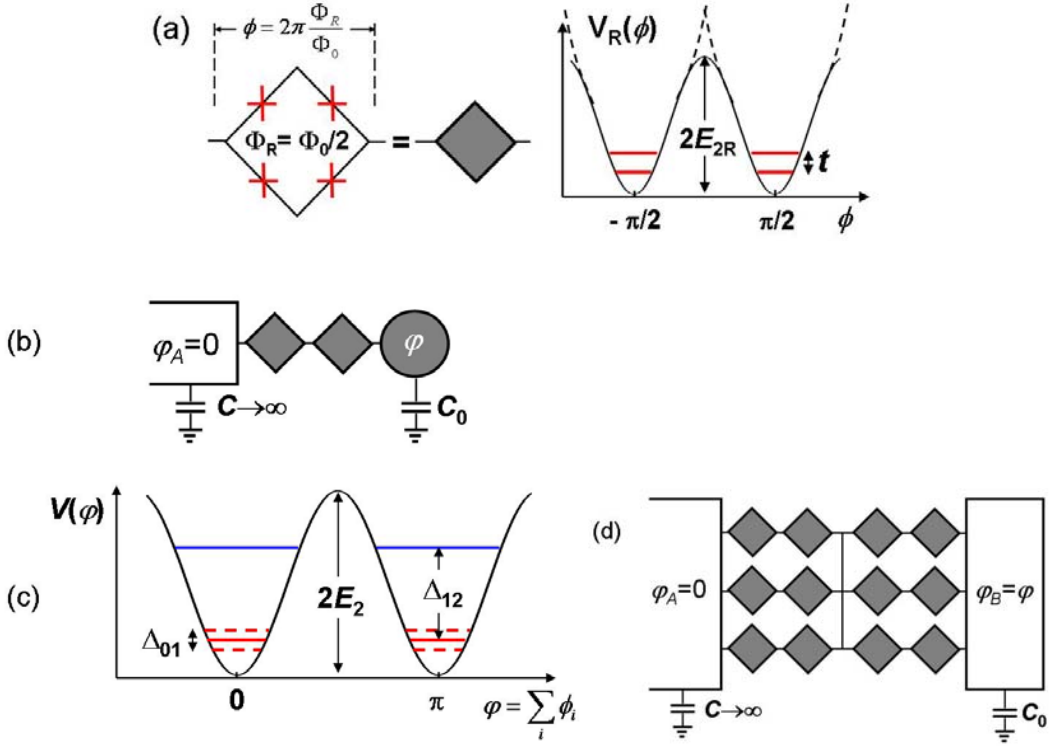

**Figure 1. Protected qubit based on "$\cos 2\phi$" Josephson elements.**

(a) The left panel shows the building block of the protected qubit (a "faulty" physical qubit): a $\cos(2\phi)$ Josephson element (a "rhombus") implemented as a superconducting loop interrupted by four nanoscale JJs (red crosses) and threaded by the magnetic flux $\Phi_R = \Phi_0/2$. The right panel shows the Josephson energy of this physical qubit, $V_R = E_{2R}\cos 2\phi$, which is doubly periodic in the phase difference $\phi$ across the rhombus. The dashed lines show $V_R(\phi)$ in the classical limit ($E_C \to 0$). Quantum fluctuations "smear" the cusps at $\phi = n\pi$ and result in tunnelling between the states $\left|\frac{\pi}{2}\right\rangle$ and $\left|-\frac{\pi}{2}\right\rangle$.

(b) A chain of two $\cos 2\phi$ elements connects an "island" with the superconducting phase $\varphi$ to a large superconducting lead with phase $\varphi_A = 0$. The effective Josephson energy of an individual rhombus, $V_R = E_{2R}\cos 2\phi$, and the island capacitance to ground, $C_0$, are chosen so that the quantum fluctuations of the phase difference $\varphi - \varphi_A = \sum_i \phi_i$ are small.

(c) The effective potential of the chain, $V(\varphi) = -E_2 \cos 2\varphi$, with two degenerate classical states at $\varphi - \varphi_A = 0, \pi$ shown by solid red lines. A finite value of $C_0$ leads to tunnelling between these states and a small level splitting $\Delta_{01}$. Higher levels are separated from this (almost) degenerate doublet by a gap $\Delta_{12}$.

(d) Connection of several rhombi chains in parallel helps to increase the depth of the effective potential $V(\varphi)$ and to suppress the transitions between qubit's logical states. The qubit logical variable is the phase of the rightmost island, $\varphi_B = \varphi$, which for a four-rhombi chain acquires values 0 or $\pi$.

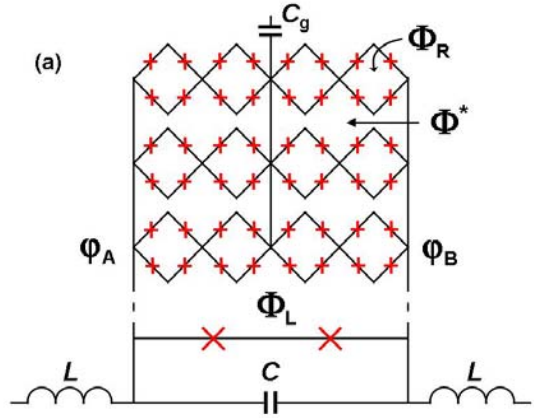

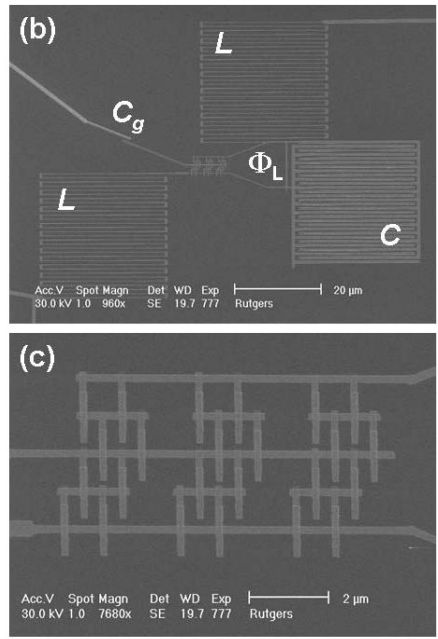

**Figure 2. The prototype of a superconducting qubit protected from local sources of noise.** Panels (a), (b), and (c) show the schematic design and the micrographs of the device, respectively. The magnetic flux $\Phi_R$ through each rhombus of an area of 1 $\mu m^2$ controls the effective Josephson energy of the rhombi. In order to probe $V(\varphi_{AB})$, three rhombi chains are included in a superconducting loop with two larger JJs (bigger red crosses on Panel (a)). The Josephson junctions are formed at each intersection of aluminum strips on Panels (c) and (d). This SQUID-like device is protected from external high-frequency noise and non-equilibrium quasiparticles generated outside of the device by two meander-type inductances $L$. To ensure the "classical" behavior of larger JJs, the SQUID-like device is shunted by an inter-digital capacitor $C \sim 6 \cdot 10^{-14}$ F.

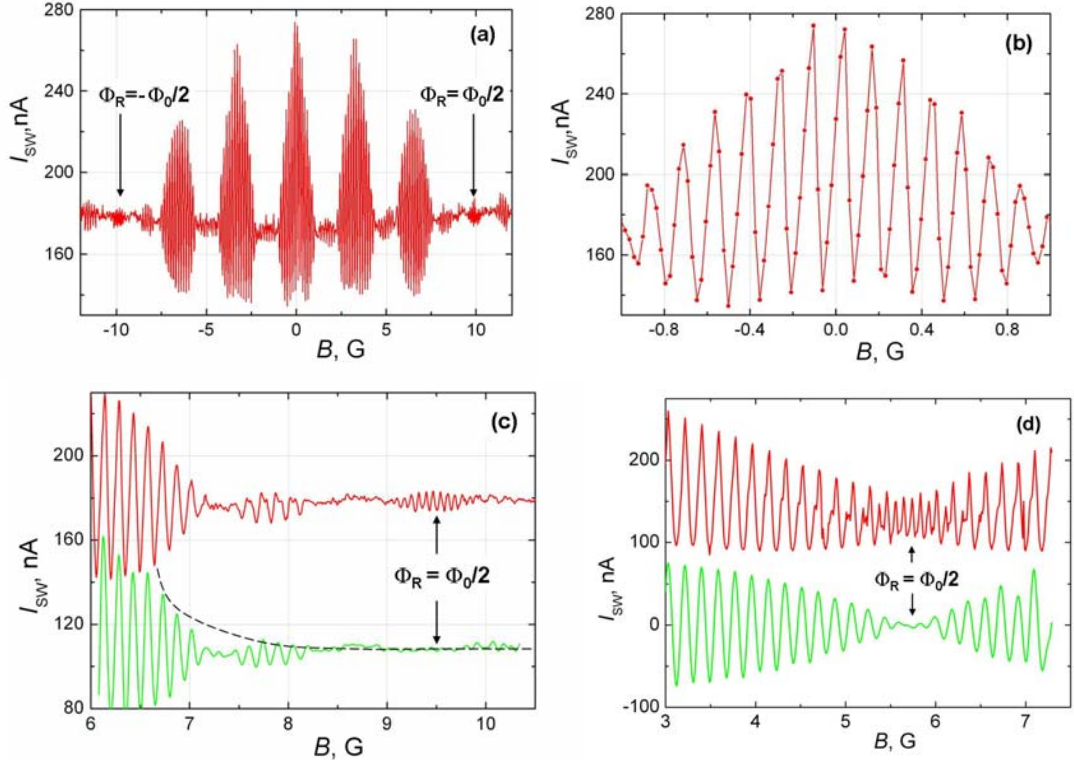

**Figure 3. Coherent transport of pairs of Cooper pairs.**

Panels (a)-(c) show oscillations of the switching current as a function of magnetic field measured with 1-ms-long current pulses for device 2 at $T = 50$ mK. Panel (a) shows these oscillations over the field range which corresponds to the magnetic flux through a single rhombus, $\Phi_R$, ranging from $-\Phi_0/2$ to $\Phi_0/2$. For almost all values of $\Phi_R$ except for $\Phi_R \approx \pm \Phi_0/2$, $I_{SW}$ oscillates with the period $\Delta\Phi_L = \Phi_0$ ($\Phi_L$ is the flux through the loop of the SQUID-type device with an area of ~110 μm$^2$) [panel (b)]. The period of oscillations is cut in half when $\Phi_R \approx \pm \Phi_0/2$ [panel (c)]. In the latter regime, the oscillations of $I_{SW}$ with the period $\Delta\Phi_L = \Phi_0/2$ are due to the correlated transport of pairs of Cooper pairs with charge 4e. The oscillations with the period of $\Delta\Phi_L = \Phi_0$ are shown as the green curve in Panel (c) (shifted for clarity down by 70 nA). Their amplitude (schematically shown as the dashed line) is strongly reduced over a relatively wide range of magnetic fields around $\Phi_R = \Phi_0/2$. For comparison, panel (d) shows the experimental data (red) and the harmonic of oscillations with the period of $\Delta\Phi_L = \Phi_0$ (green, shifted for clarity down by 130 nA) for a single two- rhombi chain. In the latter case, the suppression of the first harmonic is observed over a much narrower range of magnetic fields, in agreement with theoretical predictions.

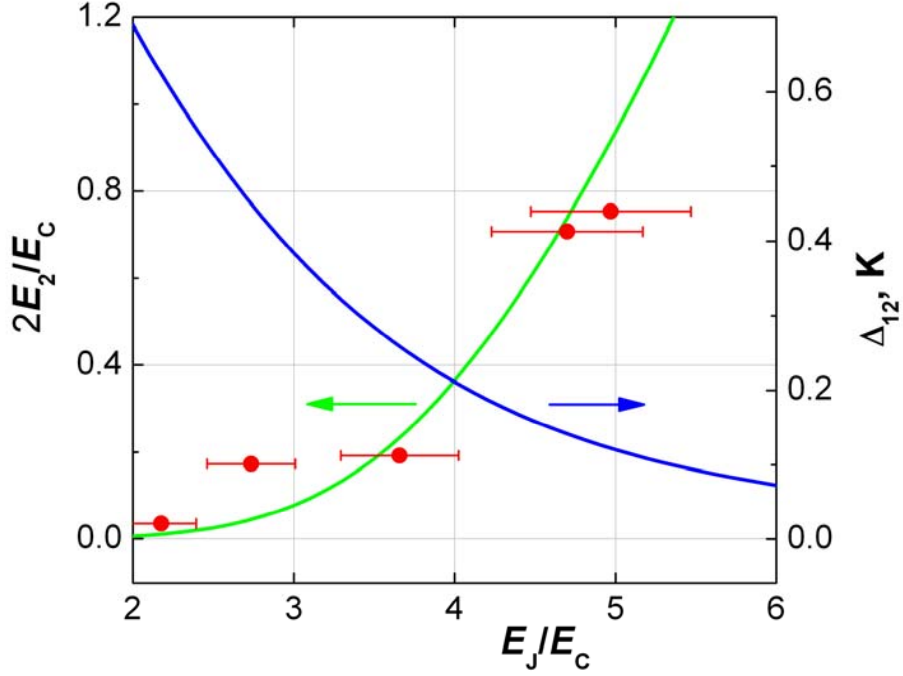

**Figure 4. Characteristic energies $2E_2$ and $\Delta_{12}$ for the devices with different values of $E_J/E_C$.**
The experimental points show the potential barrier $2E_2$ between the states $|0\rangle$ and $|\pi\rangle$ of rhombi arrays calculated from the measured amplitude of the oscillations of switching current with period $\Delta\Phi_L = \Phi_0/2$, $I_2 = \dfrac{4eE_2}{\hbar}$ (for the parameters of individual JJs in the studied devices see Table 1). The Josephson energy $E_J$ for individual JJs has been determined from $R_N$ using Ambegaokar-Baratoff formula, the Coulomb energy was estimated from the area of the junctions [C = (area)× 50 fF/µm$^2$ [20]]. The dependences $2E_2/E_C$ and $\Delta_{12}$ on $E_J/E_C$ (green and blue curves, respectively) show the results of numerical calculations for a 4x3 rhombi array (for details, see Supplementary Materials).

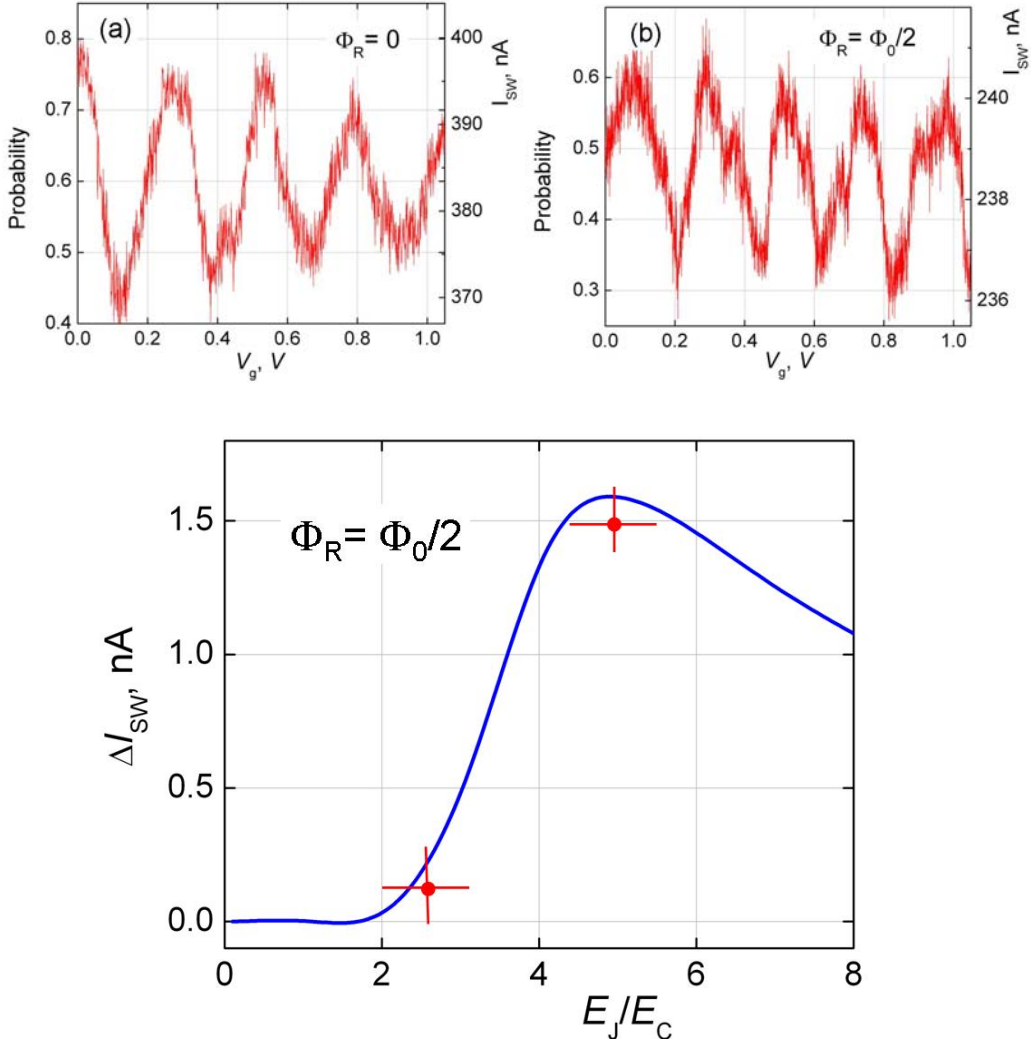

**Figure 5. Gate voltage dependence of the switching current.**

Panels (a) and (b) show the probability of switching into the resistive state for the device with $E_J/E_C \sim 4.7$, measured with a fixed amplitude of current pulses. This probability can be directly translated into the value of the switching current shown on the right vertical axes.. Panel (c) shows that the amplitude of modulation of the switching current, $\Delta I_{SW}$, measured for two devices with different values of $E_J/E_C$ in the regime $\Phi_R = \Phi/2$, is in good agreement with the result of numerical calculation (the blue curve).

*Supplementary Information*

**Device Fabrication.**

Realization of topologically protected superconducting qubits requires fabrication of nanoscale Josephson junctions (JJs) with relatively narrow margins of parameters. For an efficient protection, the values of $E_J/E_C$ for all JJs that form "rhombi" should be within ~30%, which implies that the scattering of widths of Al strips forming these JJs, $W \sim 0.15\,\mu m$, should not exceed ~10% ($E_J/E_C \sim W^4$). To reduce scattering of parameters of nanoscale Josephson junctions, we have used the so-called "Manhattan-pattern" double-layer lift-off mask schematically shown in Suppl. Fig. 1. The pattern consists of "avenues" and "streets" intersecting at right angles, the JJs are formed at each intersection of Al strips. The fabrication process consists of several steps. After fabrication of the lift-off mask on a Si substrate covered by a ~ 0.2μm-thick layer of $SiO_2$ or $Si_3N_4$, the substrate is placed in an oil-free deposition system with a base pressure ~$1\times10^{-8}$ mbar. The rotatable substrate holder is positioned at an angle $45^0$ with respect to the direction of e-gun deposition of Al. Initially, the bottom electrodes with thickness ~ 17 nm are formed by depositing Al along the direction of "avenues". Note that during this deposition, no aluminum is deposited in the "streets" because the mask thickness (~ 0.4 μm) is greater than the width of the "streets" (~ 0.15 μm). The surface of bottom electrodes is oxidized in a reduced atmosphere (~ 40 mtorr) of dry oxygen without removing the sample from the vacuum chamber. Next, the substrate holder is rotated by $90^0$, and the top electrodes with a total thickness of ~ 35 nm are deposited in two steps along the "streets" (no aluminum is deposited in the "avenues"). Depositions 2 and 3 shown in Suppl. Fig. 1 are required for a better step coverage. Finally, the sample is removed from the vacuum chamber and the lift-off mask is dissolved in the resist remover.

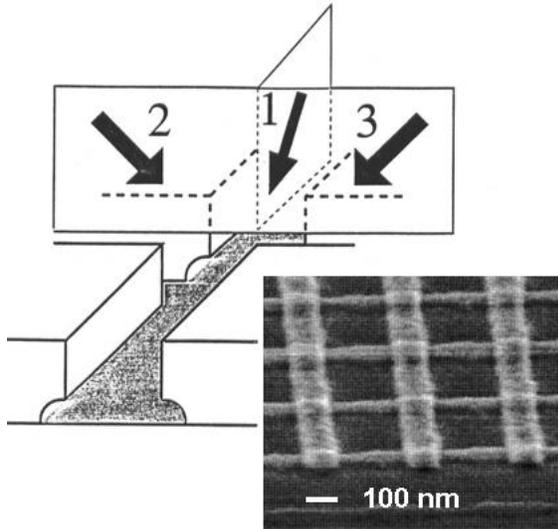

**Suppl. Fig. 1**. Schematic representation of the multi-layer deposition through an e-beam patterned mask (the so-called Manhattan pattern). The bilayer mask is formed by an e-beam resist (top layer) and copolymer (bottom layer). By an e-beam deposition of aluminum at different angles, two sets of overlapping strips are formed. The bottom film is oxidized in a reduced oxygen atmosphere prior to deposition of the second film, and the tunnel barriers are formed between the films. The microphotograph shows a test pattern formed by overlapping ~100-nm-wide Al strips.

**Measurements of the switching current.**

The current-voltage characteristics of the studied underdamped JJs are hysteretic (see, e.g., [24]): when the current $I$ exceeds the critical current $I_{C0}$, the voltage across the junction jumps up to ~ $2\Delta/e$ ~ 0.4 mV, and the junction remains in the resistive state until $I$ is reduced down to $I_r \ll I_{C0}$. For the characterization of the probability of switching into the resistive state, $P$, we have used 1-ms current pulses which were repeated every 5ms. The probability $P$ was calculated as a ratio of the number of voltage pulses $V \sim 2\Delta/e$ recorded by a pulse counter to the total number of current pulses applied to the device. Suppl. Fig. 2a shows the probability as a function of the pulse amplitude, $I$, for a fixed magnetic field $B$ which approximately corresponds to $\Phi_R = \Phi_0$ (in this regime, the effective Josephson energy of rhombi is large). The dependence of the switching current $I_{SW} = I(P=0.5)$ on $B$, shown on Suppl. Fig. 2c, has a characteristic "saw-tooth" shape (see, e.g., [24]). The width of the "probability-vs-current" curve, $\Delta I = I(P=0.9) - I(P=0.1)$, is greatly increased on the steeper slopes of this saw-tooth dependence (Fig. 2b). In this regime [$\Phi_R \approx n\Phi_0$, $\Phi_L \approx (n+1/2)\Phi_0$], the device is sensitive to the gate voltage (see Fig. 5), which reflects the fact the system can be easily excited over the barrier by the quantum fluctuations. When $\Phi_R/\Phi_0$ is half-integer, the quantum fluctuations are important for all values of $\Phi_L/\Phi_0$. This leads to an almost B-independent width of the "probability-vs-current" curve is almost $B$-independent and to the oscillatory dependence $I_{SW}(V_g)$ observed for all values of $\Phi_L/\Phi_0$.

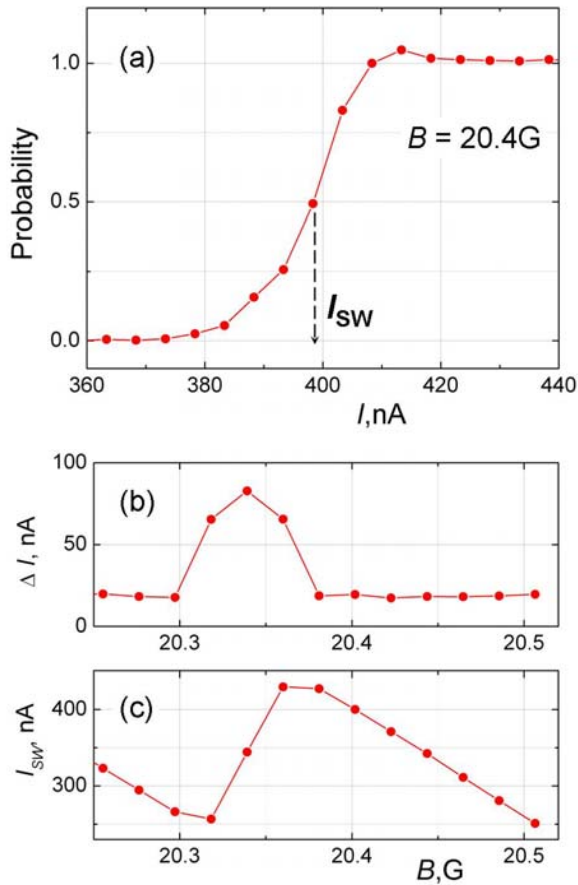

Suppl. Fig. 2. (a) The probability of switching of device 4 in the resistive state measured at $T = 50$ mK and $B = 20.4$G ($\Phi_R \cong \Phi_0$). (b) and (c) The magnetic field dependences of the width of the "probability-vs-current" curve, $\Delta I = I(P=0.9)-I(P=0.1)$, and the switching current, $I_{SW} \equiv I(P=0.5)$, respectively, measured at $T = 50$ mK.

**Theoretical estimates.**

The Josephson circuits with $E_C \leq E_J \leq \Delta$ ($\Delta$ is the superconducting order parameter) are reasonably well described by the Hamiltonian that includes only superconducting phase and conjugated to it charge degree of freedom:

$$H = -\frac{1}{2}\sum_{ij} J_{ij} \cos(\varphi_i - \varphi_j - \phi_{ij}) + 4E_C \sum_{ij} \hat{c}_{ij}^{-1} n_i n_j$$

where $J_{ij}$ is the matrix of Josephson couplings between the islands, $\phi_{ij}$ is the phase difference induced by the external magnetic field, $E_C$ is the charging energy of an individual JJ, $n_i$, $\varphi_i$ are the charge of the i[th] islands in the units of 2e and its phase, respectively, $\hat{c}$ is the capacitance matrix expressed in the units of the individual JJ capacitance. Operators $n_i$ and $\varphi_i$ are canonically conjugate to each other. For small $E_J / E_C \leq 1$, the model can be solved analytically by constructing the perturbation expansion in this parameter. For large $E_J / E_C \geq 10$, one can use a classical or a semiclassical approximations [19]. However, as explained in the text, in the experimentally relevant regime of $E_J = (3-8)E_C$, one should employ numerical calculations. The analytical methods mentioned above provide, however, a useful check on the numerical results. For the numerical estimates we use the charge representation of the model

$$H = -\frac{1}{2}\sum_i J_{i0}(b_i e^{i\phi_{i0}} + b_i^\dagger e^{-i\phi_{i0}}) - \frac{1}{4}\sum_{ij} J_{ij}(b_i b_j^\dagger e^{i\phi_{ij}} + b_j b_i^\dagger e^{-i\phi_{ij}}) + 4E_C \sum_{ij} \hat{c}_{ij}^{-1} n_i n_j$$

where $b_i$ is the charge raising operator and we have written explicitly the term describing the coupling of the islands to the leads with fixed phases $\phi_{i0}$. Direct diagonalization of small systems (consisting of up to four rhombi connected in series) by Lanczos method shows that it is sufficient to keep ~7-11 charging states on each island to get the energies of the lowest states with better than 10% accuracy for $E_J < 10 E_C$. Applying such direct diagonalization procedure to the chain of two rhombi connected to two leads with phase difference $\varphi$, we get the effective energy of this small two rhombi chain $V_{2R} = \tilde{E}_{2R} \cos(2\varphi)$ as a function of $E_J / E_C$. Because direct diagonalization of the full 12 rhombi circuit is impossible even if we keep only 3 charging states per each island, we had to use the approximate methods. The simplest of these approximation amounts to the replacement of the two rhombi chain by the effective element with $V_{2R} = \tilde{E}_{2R} \cos(2\varphi)$ Josephson energy and effective charging energy $\tilde{E}_{2C}$. The former can be found directly from the solution of the two rhombi chain with fixed phase difference across the chain while the latter can be found from the spectrum of the open two rhombi chain with the additional capacitance attached to the last island. Alternative approximation involves the exact diagonalization of the four rhombi chain with the additional capacitance attached to the middle island that emulates the effect of the other chains. We have checked that different approximations give the result for $E_2$ of the 12 rhombi circuit consistent with each other within 10%. The details of these calculations can be found elsewhere [19].

**Acknowledgements**

We thank Javier Sanchez and Sergei Pereverzev for help with experiments and O. Buisson, W. Guichard, M. Feigelman, B. Pannetier, and V. Schmidt for stimulating discussions. The work at Rutgers University was supported in part by the NSF grant ECS-0608842 and the Rutgers Academic Excellence Fund.